\documentclass[preprint,12pt]{elsarticle}
\usepackage{graphicx}
\usepackage{dcolumn}
\usepackage{amsmath}
\def\Journal#1#2#3#4{{#1} {#2} (#4) #3 }
\def\NPA{{\rm Nucl. Phys.} A}
\def\NPB{{\rm Nucl. Phys.} B}
\def\PLB{{\rm Phys. Lett.}  B}
\def\PRL{\rm Phys. Rev. Lett.}
\def\PRD{{\rm Phys. Rev.} D}
\def\PRC{{\rm Phys. Rev.} C}

\def\JPG{{\rm J. Phys.} G}

\def\ep{\epsilon}
\def\vep{\varepsilon}
\def\la{\langle}
\def\ra{\rangle}

\def\be{\begin{equation}}
\def\ee{\end{equation}}
\def\bea{\begin{eqnarray}}
\def\eea{\end{eqnarray}}
\journal{Nuclear Physics A}
\begin{document}

\begin{frontmatter}
%\vspace{2.in}
\title{Light-front dynamic analysis of transition form factors in
the process of  $P\to V\ell\nu_{\ell}$ }
\author{Ho-Meoyng Choi}
\ead{homyoung@knu.ac.kr}
\address{ Department of Physics, Teachers College, Kyungpook National University,
     Daegu, Korea 702-701}

\author{Chueng-Ryong Ji}
\ead{crji@ncsu.edu}
\address{ Department of Physics, North Carolina State University,
Raleigh, North Carolina 27695-8202, USA}

\begin{abstract}
We investigate the light-front zero-mode contribution to the weak transition
form factors between pseudoscalar and vector mesons using a covariant fermion
field theory model in $(3+1)$ dimensions. In particular, we
discuss the form factors $a_-(q^2)$ and $f(q^2)$ which have been suspected
to have the zero-mode contribution in the $q^+=0$ frame.
While the zero-mode contribution in principle depends on the form of the
vector meson vertex $\Gamma^\mu=\gamma^\mu - (2k-P_V)^\mu/D$, the form factor
$f(q^2)$ is found to be free from the zero mode if the denominator $D$ contains
the term proportional to the light-front longitudinal momentum
fraction factor $(1/x)^n$ of the struck quark with the power $n>0$. Although
the form factor $a_-(q^2)$ is not free from the zero mode,
the zero-mode contribution comes only either from the simple vertex
$\Gamma^\mu=\gamma^\mu$ term or from the other term just with a constant $D$ (i.e. $n=0$),
but not with the momentum-dependent denominator (i.e. $D\sim (1/x)^n$ with $n>0$).
We identify the zero-mode contribution to $a_-(q^2)$ and incorporate it as a convolution
of the zero-mode operator with the initial- and final-state light-front wave
functions. The covariance (i.e. frame independence) of our model has been checked
by performing the light-front calculations both in the $q^+=0$ and $q^+ > 0$
frames.
We present our numerical result of the $B\to\rho$ transition
for an explicit demonstration of our findings.
\end{abstract}
%\maketitle
%\date{}
\end{frontmatter}

\section{Introduction}
{\label{sect.I}}
The exclusive semileptonic decay processes of heavy mesons generated a
great excitement not only in extracting the most accurate values
of Cabibbo-Kobayashi-Maskawa (CKM) matrix elements but also in
testing diverse theoretical approaches to describe the internal
structure of hadrons. The great virtue of semileptonic decay processes is
that the effects of the strong interaction can be separated from
the effects of the weak interaction into a set of
Lorentz-invariant form factors, i.e., the essential informations of
the strongly interacting quark/gluon structure inside hadrons.
Thus, the theoretical problem associated with analyzing
semileptonic decay processes is essentially that of calculating the weak
form factors.

Perhaps, one of the most well-suited formulations for the analysis of exclusive processes
involving hadrons may be provided in the framework of light-front (LF) quantization~\cite{BPP}.
For its simplicity and the predictive power of the hadronic form factors in low-lying
ground-state hadrons, especially mesons, the LF
constituent quark model (LFQM) based on the LF quantization has become a very useful and popular
phenomenological tool to study various electroweak properties of
mesons~\cite{Te,Dz,CCP,CC,Ja90,CJ1,MF97,CCH,Ja99,Ja03,CHQ}.
The simplicity on the LF quantization~\cite{BPP} is mainly
attributed to the suppression of the vacuum fluctuations with the decoupling
of complicated zero modes~\cite{Zero} and the conversion of the dynamical problem from
boost to rotation.
The suppression of vacuum fluctuations is due to the
rational energy-momentum dispersion relation which correlates the signs of
the LF energy $k^-=k^0-k^3$ and the LF longitudinal momentum $k^+= k^0+k^3$.
However, the zero-mode complication in the matrix
element has been noticed for the electroweak form factors involving a
spin-1 particle~\cite{CCH,Ja99,Ja03,BCJ03,ZM05}. A growing concern
is to pin down which form factors
receive the zero-mode contributions.

The main purpose of this work is to analyze the weak form factor
$a_-(q^2)$, which has not been computed in our previous work of the
semileptonic $P\to V\ell\nu_{\ell}$ decays~\cite{ZM05}.
Unlike the form factors $g(q^2)$, $a_+(q^2)$, and $f(q^2)$, which
can be obtained from the plus component of the currents~\cite{ZM05}, one needs to use the perpendicular
(or minus) components of the currents to obtain $a_-(q^2)$. In this work, we use the
perpendicular components of the axial-vector currents with the transverse polarization
to obtain $a_-(q^2)$ and analyze the existence/absence of the zero mode.

The paper is organized as follows. In Section 2, we discuss the $P\to V\ell\nu_{\ell}$
semileptonic decays using an exactly solvable model based on the covariant
Bethe-Salpeter (BS) model of $(3+1)$-dimensional fermion field theory.
In Section 3, we present our LF calculation of the weak form factors in the $q^+ > 0$
frame and discuss the result in the $q^+\to 0$ limit for the analysis on
the existence/absence of the zero-mode contribution to the form factors.
Especially, we identify the zero-mode contribution to $a_-(q^2)$ and incorporate it as a
convolution of the zero-mode operator with the initial- and final-state LF wave
functions.
We also present our numerical result for the explicit demonstration of our findings.
Summary and discussion follow in Section 4.
In the appendices A and B, we summarize the LF results of the trace terms for the weak current
matrix element and the results of the weak form factors obtained from the $q^+> 0$ frame, respectively.

\section{Model Description}
The Lorentz-invariant transition form factors $g$, $f$, $a_{+}$, and
$a_{-}$ between a pseudoscalar meson with four-momentum $P_1$
and a vector meson with four-momentum $P_2$ and helicity $h$ are
defined by the matrix elements of the electroweak current
$J^\mu_{V-A}$ from the
initial-state $|P_1;00\ra$ to the final-state $|P_2;1h\ra$~\cite{AW}:
\begin{eqnarray}
\la P_2;1h|J^\mu_{V-A}|P_1;00\ra
 &=&
i g(q^2) \varepsilon^{\mu\nu\alpha\beta}
\epsilon^*_{\nu}P_\alpha q_\beta -f(q^2) \epsilon^{*\mu}
\nonumber\\
&& -a_{+}(q^2)(\epsilon^{*}\cdot P)
P^\mu -a_{-}(q^2)(\epsilon^{*}\cdot P)q^\mu,
{\label{eq:1}}
\end{eqnarray}
where $P=P_1 + P_2$ and $q = P_1 - P_2$ is the four-momentum transfer
to the lepton pair ($\ell\nu_\ell$).
The polarization vector $\epsilon^*=\epsilon^*(P_2,h)$ of the
final-state vector
meson satisfies the Lorentz condition $\epsilon^*\cdot P_2 = 0$.
The polarization vectors used in this analysis are given by
\bea{\label{pol_vec}}
\epsilon^\mu(\pm 1)&=&[\ep^+,\ep^-,\ep_\perp]
=\biggl[0,\frac{2}{P^+_2}{\bf\epsilon}_\perp(\pm)\cdot{\bf P_{2\perp}},
{\bf\epsilon}_\perp(\pm 1)\biggr],
\nonumber\\
{\bf\epsilon}_\perp(\pm 1)&=&\mp\frac{(1,\pm i)}{\sqrt{2}},\;
\epsilon^\mu(0)=
\frac{1}{M_2}\biggl[P^+_2,\frac{{\bf P}^2_{2\perp}-M^2_2}{P^+_2},
{\bf P}_{2\perp}\biggr].
\eea

While the form factor $g(q^2)$ is associated with the vector
current $V^\mu$, the rest of the form factors $f(q^2)$, $a_{+}(q^2)$, and
$a_{-}(q^2)$ are coming from the axial-vector current $A^\mu$.

The transition form factors defined in Eq.~(\ref{eq:1}) are often
given by the Bauer, Stech, and Wirbel (BSW) convention~\cite{BSW},
\bea
V(q^2) &=& -(M_1 + M_2)g(q^2),\nonumber\\
A_1(q^2) &=& -{f(q^2) \over {M_1 + M_2}}, \nonumber \\
A_2(q^2) &=& (M_1 + M_2)a_{+}(q^2), \nonumber\\
A_0(q^2) &=& \frac{-1}{2 M_2} \biggl[f(q^2)+ q\cdot P a_{+}(q^2)
+ q^2 a_{-}(q^2) \biggr],
{\label{BSWf}}
\eea
where $M_1$ and $M_2$ are the physical pseudoscalar and vector meson
masses, respectively, and $q\cdot P=M^2_1-M^2_2$.

\begin{figure}
\begin{center}
\includegraphics[width=3.0in]{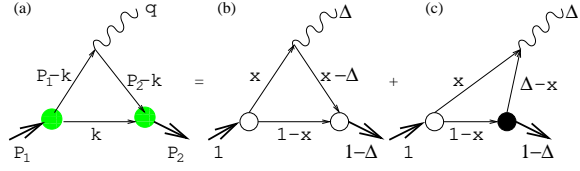}
\end{center}
\caption{The covariant diagram (a) corresponds to the sum of the
LF valence diagram (b) and the nonvalence diagram (c). The large white
and black blobs at the meson-quark vertices in (b) and (c) represent
the ordinary LF wave function and the nonvalence wave function vertices,
respectively.}
\label{fig1}
\end{figure}

The exactly solvable model based on the covariant BS model of $(3+1)$-dimensional
fermion field theory~\cite{MF97,BCJ03} enables us to derive the transition form
factors between pseudoscalar and vector mesons explicitly.
The covariant diagram shown in Fig. 1(a) is in general equivalent to the sum of the LF
valence diagram [Fig. 1(b)] and the nonvalence diagram [Fig. 1(c)].
The matrix element
$\la J^\mu_{V-A}\ra_h\equiv\la P_2;1h|J^\mu_{V-A}|P_1;00\ra$ obtained from the
covariant diagram of Fig. 1(a) is given by
\begin{equation}{\label{eq:4}}
\la J^\mu_{V-A}\ra_{h}
=  ig_1 g_2\Lambda^2_1\Lambda^2_2
\int\frac{d^4k}{(2\pi)^4} \frac{(S^\mu_{V-A})_h} {N_{\Lambda_1}
N_{1} N_{\bar q} N_{2} N_{\Lambda_2}},
\end{equation}
where $g_{1(2)}$ is the normalization factor  which can be fixed by
requiring charge form factor of pseudoscalar (vector) meson to be
unity at $q^2=0$. To regularize the covariant fermion
triangle loop in $(3+1)$ dimensions, we replace the
point gauge-boson vertex $\gamma^\mu (1-\gamma_5)$ by a non-local
(smeared) gauge-boson vertex
$({\Lambda_1}^2 / N_{\Lambda_1})\gamma^\mu (1-\gamma_5)
( {\Lambda_2}^2 / N_{\Lambda_2})$, where $N_{\Lambda_{1(2)}}
=p_{1(2)}^2-{\Lambda_{1(2)}}^2+i\ep$, and $\Lambda_1$
and $\Lambda_2$ play the role of momentum cut-offs similar to the Pauli-Villars
regularization.  The rest of the
denominators in Eq.~(\ref{eq:4}) coming from the intermediate fermion
propagators in Fig.~\ref{fig1}(a) are given by
\be
N_1 = p_1^2 -{m_1}^2 + i\vep,\;
N_{\bar q} = k^2 - m^2 + i\vep, \;
N_{2} = p_2^2 -{m_2}^2 + i\vep,
{\label{eq:5}}
\ee
where $m_{1(2)}$ and $m$ are the masses of the constituents
carrying the intermediate four-momenta $p_{1(2)}=P_{1(2)} -k$ and $k$, respectively.

The trace term $(S^\mu_{V-A})_h$ in Eq.~(\ref{eq:4}) is given by
\be
(S^\mu_{V-A})_h = {\rm Tr}[(\not\!p_2 + m_2)\gamma^\mu (1-\gamma_5)
(\not\!p_1 +m_1)\gamma_5(-\not\!k + m)\epsilon^*\cdot\Gamma],
{\label{eqq:4}}
\ee
where the initial-state pseudoscalar meson vertex operator is $\gamma^5$ and
the final-state vector meson vertex operator $\Gamma^\mu$
is given by
\begin{equation}\label{eq5}
\Gamma^\mu=\gamma^\mu-\frac{(P_2-2k)^\mu}{D}.
\end{equation}
While $\gamma^\mu$ is intrinsic to the vector meson vertex, the model-dependence of
vector meson is implemented through the factor $D$ in Eq.~(\ref{eq5}).
Frequently used $D$ factor is either constant ($D_{\rm con}$) or covariant ($D_{\rm cov}$):
\bea\label{D-term}
&&(1)\; D_{\rm con}=M_2 + m_2 + m ,
\nonumber\\
&&(2) \; D_{\rm cov}
=\frac{2k\cdot P_2 + M_2(m_2 + m) -i\epsilon}{M_2}.
\eea
We note that the $D$ factor behaves like $(1/x)^n$ as the LF longitudinal momentum fraction
$x$ goes to zero (i.e. $x\to 0$), where $n=0$ and $1$ for the cases of (1) and (2) of Eq.~(\ref{D-term}),
respectively.

As discussed in our previous work~\cite{ZM05}, Jaus's prescription~\cite{Ja99}
to find the zero mode is limited to the case of $D=D_{\rm con}$.
In this work, we analyze $a_-(q^2)$ for both cases of (1) and (2) of Eq.~(\ref{D-term}) and
confirm again that Jaus's prescription applies only to the case (1) but not to the case (2).
We also apply the LF version of the $D$ factor,  i.e.
$D_{\rm LF}=M_0+m_2 + m$ with the invariant mass $M_0$ of the vector meson.
This case corresponds to $n=1/2$ and Jaus's prescription doesn't apply to this case either
as we will discuss in the next section.

\section{Light-front calculation of the weak form factors}
In the $q^+>0$ frame, the covariant diagram Fig.~\ref{fig1}(a) corresponds to the sum of the
LF valence diagram (b) defined in $0<k^+<P^+_2$ region and the
nonvalence diagram (c) defined in $P^+_2<k^+<P^+_1$ region. The large white
and black blobs at the meson-quark vertices in (b) and (c) represent
the ordinary LF wave functions and the non-wave-function vertex~\cite{BCJ03},
respectively.

Defining $\Delta =q^+/P^+_1$ and the longitudinal momentum fraction factor
$x=p^+_1/P^+_1$ ($1-x=k^+/P^+_1$) for the struck (spectator) quark,
we should note that the nonvalence region (i.e. $0<x<\Delta$) of integration shrinks
to the end point $x=0$ in the $q^+\to 0$ (i.e. $\Delta\to 0$) limit.
The virtue of taking $q^+=0$ frame is to obtain the form factor by calculating only the
valence diagram (i.e. $0 < x <1$) because the nonvalence diagram does not contribute
if the integrand is free from the singularity in $p^-_1\sim 1/x$.
However, if the integrand has a singularity as $x\to 0$, then one should take into
account not only the valence diagram but also the nonvalence diagram
because the latter can also give nonvanishing
contribution even if the integration range of this diagram shrinks to the end point $x=0$.
Thus, one needs to analyze carefully if the contribution from the nonvalence diagram in the
$q^+=0$ frame occurs or not, in order to correctly utilize the $q^+=0$ frame without any error.
Calling such contributions from the end point $x=0$ as zero modes,
we investigate them for the form factors in the $P\to V\ell\nu_\ell$ transition.

In order to check the existence/absence of
the zero-mode contribution to the hadronic matrix element given by Eq.~(\ref{eq:4}),
we first choose $q^+ >0$ frame and then take $q^+\to 0$ limit.
Our analysis for the zero mode is based on the
$q^+=0$ [or Drell-Yan(DY)] frame~\cite{DYW}:
\bea\label{DYW}
P_1&=& (P^+_1, P^-_1,{\bf P}_{1\perp})=(P^+_1,\frac{M^2_1}{P^+_1},{\bf 0}_\perp),
\;
P_2= (P^+_1,\frac{M^2_2-q^2}{P^+_1},-{\bf q}_\perp),
\nonumber\\
q &=& (0, \frac{M^2_1-M^2_2+q^2}{P^+_1},{\bf q}_\perp),
\eea
where $q^2 = -{\bf q}^2_\perp$ is the spacelike gauge boson momentum transfer.
The weak form factors in the timelike $q^2$ region can be obtained by the analytic
continuation from the spacelike $q^2$ region.
%The relevant quark momentum variables in the $q^+=0$ frame are
%$p^+_1= x P^+_1$, $k^+=(1-x)P^+_1$, and ${\bf p}_{1\perp}=-{\bf k}_\perp$.

The relations between the current matrix
elements and the weak form factors in this $q^+=0$ frame are as follows:
\bea{\label{ch:4}}
g^{\rm DY}(q^2)&=& -\frac{\sqrt{2}q^R}{q^2}\la J^+_V\ra_{h=1},
\;
a^{\rm DY}_+(q^2) = -\frac{q^R}{q^2\sqrt{2}}\la J^+_A\ra_{h=1},
\nonumber\\
 f^{\rm DY}(q^2)&=& (q^2-q\cdot P)a^{\rm DY}_+(q^2)
+ M_2\la J^+_A\ra_{h=0},
\nonumber\\
a^{\rm DY}_-(q^2)&=& a^{\rm DY}_+(q^2) - \frac{1}{q^2}[ f^{\rm DY}(q^2)
+ \frac{\sqrt{2}}{q^L}\la J^\perp_A\cdot{\bf q}_\perp\ra_{h=1}],
\eea
where $q^{R(L)}=q_x \pm iq_y$. Since the form factors ($g,a_+, f$) have been analyzed
in our previous work~\cite{ZM05}, we focus on the calculation of the form factor
$a^{\rm DY}_-(q^2)$ (i.e. $\la J^\perp_A\ra_{h=1}$) to find if the zero mode exists
or not in this work.

\subsection{Valence contribution}
In the valence region $\Delta < x <1$, the pole $k^-=k^-_{\rm on}=({\bf
k}^2_\perp + m^2_{\bar q} -i\ep)/k^+$ (i.e., the spectator quark) is located in the lower
half plane of
the complex $k^-$-variable.  Thus, the Cauchy integration formula for the
$k^-$ integral in Eq.~(\ref{eq:4}) gives
  \be\label{va1}
 \la J^\mu_{V(A)}\ra_h =
\frac{N}{16\pi^3}\int^1_0 \frac{dx}{(1-x)}\int d^2{\bf k}_\perp
\chi_1(x,{\bf k}_\perp)[S^\mu_{V(A)}]_h\chi_2 (x, {\bf k'}_\perp),
  \ee
where $N=g_1g_2\Lambda^2_1\Lambda^2_2$. The LF vertex functions
$\chi_1$ and $\chi_2$ are given by
 \bea\label{va2}
 \chi_{1(2)}(x,{\bf k}^{(\prime)}_\perp) =
 \frac{1}{ x^2 (M^2_{1(2)} - M^{(\prime)2}_0)(M^2_{1(2)}-M^{(\prime)2}_{\Lambda_{1(2)}})},
\eea
where ${\bf k'}_\perp={\bf k}_\perp + (1-x) {\bf q}_\perp$ and
 \be\label{va3}
M^{(\prime)2}_0 =
\frac{{\bf k}^{(\prime)2}_\perp + m^2}{1-x}
+ \frac{{\bf k}^{(\prime)2}_\perp + m^2_{1(2)}}{x},
\;
M^{(\prime)2}_{\Lambda_{1(2)}}= M^{(\prime)2}_0(m_{1(2)}\to\Lambda_{1(2)}).
 \ee

In our trace term $[S^\mu_{V(A)}]_h$ calculation , we separate Eq.~(\ref{eqq:4})
into the on-mass-shell propagating part $[S^\mu_{V(A)}]^{\rm on}_h$ and
the off-mass-shell instantaneous part $[S^\mu_{V(A)}]^{\rm inst}_h$, i.e.
\be{\label{sep}}
[S^\mu_{V(A)}]_h = [S^\mu_{V(A)}]^{\rm on}_h + [S^\mu_{V(A)}]^{\rm inst}_h,
\ee
via
\be{\label{ID}}
\not\!p + m =
(\not\!p_{\rm on} + m) + \frac{1}{2}\gamma^+(p^- - p^-_{\rm on}).
\ee
While the on-mass-shell part indicates that all
three quarks are on their respective mass shell, i.e. $k^-=k^-_{\rm on}$
and $p^-_{i}=p^-_{i\rm on}(i=1,2)$, the instantaneous part
includes the term proportional
to $\delta p_i^-=p^-_i-p_{i\rm on}^-(i=1,2)$ and
$\delta k^-=(k^- - k^-_{\rm on})$~\cite{BCJ03}.
The explicit forms of $[S^\mu_{V(A)}]_h$
in Eq.~(\ref{eqq:4}) are presented in the appendix A.

\subsection{Zero-mode contribution }
In the nonvalence region $0<x<\Delta$, the poles are at $p^-_1=p^-_{1\rm
on}(m_1) = (m^2_1 +{\bf k}^2_\perp -i\ep)/p^+_1$ (from the struck quark propagator)
and $p^-_1=p^-_{1\rm on}(\Lambda_1) = (\Lambda^2_1+{\bf k}^2_\perp -i\ep)/
p^+_1$ (from the smeared quark-gauge-boson vertex), which are located in the upper
half plane of the complex $k^-$-variable.
To investigate the zero-mode contribution, we need to analyze the nonvalence diagram
[Fig.~\ref{fig1}(c)] in the $\Delta\to 0$ limit, where the nonvalence region shrinks
to the end point $x=0$.

To handle the complexity of treating double $p^-_1$-poles from $N_{\Lambda_1}$
and $N_1$, we decompose the product of five denominators given in Eq.~(\ref{eq:4})
into a sum of terms containing three propagators as
follows:
\be\label{decom}
  \frac{1}{N_{\Lambda_1} N_{1} N_{\bar q} N_{2} N_{\Lambda_2}}
= \frac{1}{({\Lambda_1}^2-{m_1}^2)({\Lambda_2}^2-{m_2}^2)}
 \frac{1}{N_{\bar q}}\biggl( \frac{1}{N_{\Lambda_1}} -
\frac{1}{N_{1}}\biggr) \biggl(\frac{1}{N_{\Lambda_2}} - \frac{1}{N_{2}} \biggr).
\ee
From this decomposition, one may have zero-mode contribution proportional to $\delta(x)$
from the $p^-_1$-pole~\footnote{Note that $p^-_2$ and $-k^-$ show the same singular
behavior as $p^-_1$, i.e. $p^-_1(=p^-_2=-k^-)\sim 1/x$ as $x\to 0$.} (if exists) in the numerator.
For instance, the $k^-$ integration of $p^-_1/N_{\bar q} N_{\Lambda_1}N_{\Lambda_2}$
having $p^-_1=p^-_{1\rm on}(\Lambda_1)$ pole (i.e. $N_{\Lambda_1}\to 0$) gives
the following nonvanishing zero-mode contribution
\be\label{ZP}
 \lim_{\Delta\to 0}\int_{nv}dk^-
 \frac{p^-_1}{N_{\bar q} N_{\Lambda_1}N_{\Lambda_2}}
  =2\pi i
 \frac{\delta(x)}{\Lambda^2_{2\perp}-\Lambda^2_{1\perp}}
 \ln\frac{\Lambda^2_{2\perp}}{\Lambda^2_{1\perp}},
 \ee
where $\Lambda^2_{i\perp}=\Lambda^2_i + {\bf p}^2_{i\perp}$.
The appearance of $\delta(x)$ in our analysis is closely related to the findings
in Ref.~\cite{Zero}. It is very important to note that such zero mode in Eq.~(\ref{ZP})
is absent if $p^-_1$ is combined with a factor $x^n$ with $n>0$, i.e.
$x^n p^-_1/(N_{\bar q} N_{\Lambda_1}N_{\Lambda_2})$. From the power counting of $x$ for the
$D$ factor used in the present analysis, one can
easily see that the nonvanishing zero-mode contribution to
$(p^-_1/D)/(N_{\bar q} N_{\Lambda_1}N_{\Lambda_2})$ exists only when
$D=D_{\rm con}$ (i.e. $n=0$), but absent when $D_{\rm cov}$ (i.e. $n=1$)
or $D_{\rm LF}=M'_0+m_2 +m$ (i.e. $n=1/2$) is used.

In our previous work~\cite{ZM05} for the calculation of the form factors
$g(q^2),a_+(q^2)$, and $f(q^2)$,
we found that only the form factor $f(q^2)$ (i.e. $\la J^+_A\ra_{h=0}$) may receive
the zero-mode contribution from the $p^-_1$ term in $(S^+_A)_{h=0}$.
From the power counting rule for $p^-_1$ (or $1/x$) in $(S^+_A)_{h=0}$,
we obtained the suspected zero-mode contribution as
$(S^+_A)^{\rm Z.M.}_{h=0}= \lim_{\Delta\to 0} (S^+_A)^{\rm inst}_{h=0}=2\epsilon^{*+}_{h=0}(p^-_1/D)$.
This contribution is nonvanishing only if
$D=D_{\rm con}$  but vanishes if $\Gamma^\mu=\gamma^\mu$ (i.e. $1/D=0$),
$D=D_{\rm cov}$ or $D_{\rm LF}$ is used~\cite{ZM05}.

To find the zero-mode contribution to $a_-(q^2)$ defined by Eq.~(\ref{ch:4}),
we need to analyze the zero-mode contribution
to $\la J^\perp_A\ra_{h=1}$ since it may come from the $p^-_1$ term in $(S^\perp_A)_{h=1}$.
From the power counting rule for $p^-_1$  in $(S^\perp_A)_{h=1}$, we find
 \be\label{ZMS}
(S^\perp_A)^{\rm Z.M.}_{h=1}=\lim_{\Delta\to 0}(S^\perp_{A})^{\rm inst}_{h=1}
 = 2 p^-_1\biggl[ (m_1+m_2)\ep^*_\perp
 + 2\frac{k_{\rm on}\cdot\ep^*}{D}(2{\bf p}_{1\perp}-{\bf q}_\perp)\biggr].
 \ee
We note that only the instantaneous part
$(S^\perp_{A})^{\rm inst}_{h=1}$ in Eq.~(\ref{SAon2}) contributes to the zero mode.
The first term in the square bracket of Eq.~(\ref{ZMS}) comes from the model-independent
intrinsic $\gamma^\mu$ part, while the second term comes from the $(2k-P_V)^\mu/D$ part.
It is important to note that
the zero-mode contribution to $(S^\perp_A)_{h=1}$ comes already from
the model-independent intrinsic $\gamma^\mu$ part in the computation of $a_-(q^2)$.
As in the case of $f(q^2)$, however, the zero-mode contribution from the $p^-_1/D$ term
is nonvanishing only if $D=D_{\rm con}$  but vanishes if
$D=D_{\rm cov}$ or $D_{\rm LF}$.

The net zero-mode contribution to $\la J^\perp_A\ra_{h=1}$ is then obtained as
$\la J^\perp_A\ra^{\rm Z.M.}_{h=1}$ =
$[J^\perp_{\Lambda_1\Lambda_2} ]_{\rm Z.M.}$
- $[J^\perp_{\Lambda_1 m_2} ]_{\rm Z.M.}$
- $[J^\perp_{m_1\Lambda_2}]_{\rm Z.M.}$
+ $[J^\perp_{m_1m_2}]_{\rm Z.M.}$ from the decomposition of
the denominators according to Eq.~(\ref{decom}).
 For instance, we define the zero mode contribution to $1/N_{\bar q} N_{\Lambda_1}
N_{\Lambda_2}$ term in Eq.~(\ref{decom}) as
\be\label{JZM}
 [J^\perp_{\Lambda_1\Lambda_2}]_{\rm Z.M.}
  =\lim_{\Delta\to 0}\int_{nv}\frac{d^4k}{(2\pi)^4}
 \frac{[S^\perp_A(p^-_1=p^-_{1\rm on}(\Lambda_1))]^{\rm Z.M.}_{h=1} }
 {N_{\bar q} N_{\Lambda_1}N_{\Lambda_2}}.
 \ee
The zero-mode contributions to the other three terms can be
defined the same way as in Eq.~(\ref{JZM}).
Therefore, as far as the $D_{\rm LF}$ or $D_{\rm cov}$ is used,
the nonvanishing zero-mode contribution to $(S^\perp_A)^{\rm Z.M.}_{h=1}$ comes only from
the intrinsic $\gamma^\mu$ part. In this case, the nonvanishing
zero-mode contribution to
$\la J^\perp_A\cdot {\bf q}_\perp\ra_{h=1}$ is given by

 \bea\label{JPZM}
 \la J^\perp_A\cdot {\bf q}_\perp\ra^{\rm Z.M.}_{h=1}
 &=& \frac{N}{8\pi^2({\Lambda_1}^2-{m_1}^2)({\Lambda_2}^2-{m_2}^2)}
\frac{q^L}{\sqrt{2}}(m_1+m_2)
\nonumber\\
&&\times\int^1_0 dz \ln\biggl( \frac{B_{\Lambda_1 m_2}
 B_{m_1\Lambda_2}} {B_{\Lambda_1\Lambda_2}B_{m_1m_2}} \biggr),
 \eea

where
 \be\label{ap:28}
B_{ab} =  (1-z)a^2 + z b^2 - z(1-z) q^2.
 \ee

\subsection{Effective inclusion of the zero-mode in the valence region }

We may identify the zero-mode operator that is convoluted with
the initial and final state valence wave functions to generate the zero-mode
contribution given by Eq.~(\ref{JPZM}).
Since our findings agree with Jaus's results for the intrinsic $\gamma^\mu$ part as well as
the model-dependent $(P_2-2k)^\mu/D$ part if the factor $D$ is constant,
our method for those parts that we agree with Jaus can also be realized effectively
by the Jaus's method~\cite{Ja99}
using the orientation of the LF plane characterized by the invariant equation
$\omega\cdot x=0$~\cite{CDKM,SCC}, where $\omega$ is an arbitrary lightlike
four vector.  While the physical amplitudes should not depend
on the orientation of the LF plane, the LF matrix elements can
acquire a spurious $\omega$ dependence. This problem is closely associated with the violation of
rotational invariance in the computation of the matrix element of a one-body current.
In order to treat the complete Lorentz structure of a hadronic matrix
element, the authors in~\cite{Ja99,CDKM} have developed a method to identify and
separate spurious ($\omega$ dependent)
contributions  to the hadronic
form factors. Below, we summarize the result of zero-mode contribution obtained
from the method by Jaus~\cite{Ja99} and discuss the equivalence to
our result of zero-mode contribution.

By adopting the $\omega$ dependent LF covariant approach as in~\cite{Ja99,CDKM},
we identify the zero-mode operator that is convoluted with the initial and final
state valence wave functions to generate the zero-mode contribution to the form
factor $a_-(q^2)$. In order to do this, we first decompose the four vector
$p^\mu_1$ in terms of $P, q$, and
$\omega$ with $\omega=(1,0,0,-1)$ as follows~\cite{Ja99}:
 \be\label{eq:j1}
p^\mu_1 = P^\mu A^{(1)}_1 + q^\mu A^{(1)}_2 + \frac{1}{\omega\cdot P}\omega^\mu C^{(1)}_1.
 \ee
The coefficients in Eq.~(\ref{eq:j1}) are given by
 \bea\label{eq:j2}
A^{(1)}_1&=& \frac{\omega\cdot p_1}{\omega\cdot P}=\frac{x}{2},
\nonumber\\
A^{(1)}_2&=& \frac{1}{q^2}\biggl( p_1\cdot q
- (q\cdot P)\frac{\omega\cdot p_1}{\omega\cdot P}\biggr)
=\frac{x}{2} + \frac{{\bf k_\perp\cdot{\bf q}_\perp}}{q^2},
\nonumber\\
C^{(1)}_1 &=& p_1\cdot P - P^2 A^{(1)}_1 - q\cdot P A^{(1)}_2 = Z_2 - N_{\bar q},
 \eea
where
 \be\label{eq:j3}
Z_2 = x(M^2_1 - M^2_0) + m^2_1 - m^2_{\bar q} + (1-2x)M^2_1
- [q^2 + q\cdot P]\frac{{\bf k_\perp\cdot{\bf q}_\perp}}{q^2}.
 \ee
Note that only the coefficient $C^{(1)}_1$ which is combined with $\omega^\mu$
depends on $p^-_1$ (i.e. zero mode). In this exactly solvable BS model,
the zero-mode contribution from $p^-_1$ is exactly opposite to that from $N_{\bar q}$~\cite{Bc09},
i.e.
 \bea\label{eq:j4}
 I[p^-_1]_{\rm Z.M.}&=&iN\int_{\rm Z.M.} \frac{d^4k}{(2\pi)^4}
\frac{p^-_1}{N_{\Lambda_1}N_1 N_{\bar q} N_2 N_{\Lambda_2}}
\nonumber\\
&=& \frac{N}{16\pi^2({\Lambda_1}^2-{m_1}^2)({\Lambda_2}^2-{m_2}^2)}
\int^1_0 dz  \ln\biggl( \frac{B_{\Lambda_1 m_2}
 B_{m_1\Lambda_2}} {B_{\Lambda_1\Lambda_2}B_{m_1m_2}} \biggr)
 \nonumber\\
 &=&-I[N_{\bar q}]_{\rm Z.M.}.
 \eea
Furthermore, the zero-mode contribution $I[N_{\bar q}]_{\rm Z.M.}$ from $N_{\bar q}$ is exactly the
same as the valence contribution $I[Z_2]_{\rm val}$ from $Z_2$, where $I[Z_2]_{\rm val}$ is given by
 \be\label{eq:j5}
I[Z_2]_{\rm val}= \frac{N}{16\pi^3}\int^1_0 \frac{dx}{1-x}\int d^2{\bf k}_\perp
 \chi_1(x,{\bf k}_\perp)\chi_2(x,{\bf k}'_\perp) Z_2.
 \ee
From the identities in Eqs.~(\ref{eq:j4}) and (\ref{eq:j5}), the replacement
$N_{\bar q}\to Z_2$(or equivalently $p^-_1\to -Z_2$) in the spurious $\omega$
dependent (i.e. the zero-mode related) term
$C^{(1)}_1$ in Eq.~(\ref{eq:j2}) makes the amplitude free of any $\omega$ dependence,
and effectively includes the zero-mode contribution from $p^-_1$ in the valence region
with the help of Eq.~(\ref{eq:j5}). Using this prescription, we can effectively include
the zero-mode contribution to
$\la J^\perp_A\cdot {\bf q}_\perp\ra_{h=1} $
in the LF valence region.
For the intrinsic $\gamma^\mu$ part, as an example,
the nonvanishing zero-mode contribution to
$\la J^\perp_A\cdot {\bf q}_\perp\ra_{h=1}$ is obtained as
\be\label{JPZM2}
\la J^\perp_A\cdot {\bf q}_\perp\ra^{\rm Z.M.}_{h=1}
= -\frac{q^L}{\sqrt{2}}
\frac{N}{8\pi^3}\int^1_0\frac{dx}{(1-x)}
\int d^2{\bf k}_\perp\; \chi_1\chi_2
(m_1+m_2)Z_2,
\ee
and the full result for
$\la J^\perp_A\cdot {\bf q}_\perp\ra_{h=1}$ is given by
$\la J^\perp_A\cdot {\bf q}_\perp\ra^{\rm full}_{h=1}
= \la J^\perp_A\cdot {\bf q}_\perp\ra^{\rm val}_{h=1}
+
\la J^\perp_A\cdot {\bf q}_\perp\ra_{h=1}^{\rm Z.M.}$.
We should note that Eq.~(\ref{JPZM}) and Eq.~(\ref{JPZM2})
coincide exactly.

\subsection{Transition Form Factors for $D=D_{\rm cov}$ and
$D_{\rm LF}$}
Since more realistic LFQM uses $D=D_{\rm cov}$ or $D_{\rm LF}$ instead of
$D_{\rm con}$, we obtain the transition form factors which are valid
when $D=D_{\rm cov}$ or $D_{\rm LF}$.
In this case, the three form factors $g(q^2)$, $a_+(q^2)$, and $f(q^2)$ can be obtained
 without encountering the zero-mode contribution as we have already
proved in Ref.~\cite{ZM05}. On the other hand, the form factor $a_-(q^2)$  receives the
zero-mode contribution from $\la J^\perp_A\cdot {\bf q}_\perp\ra^{\rm Z.M.}_{h=1}$,
which comes from the intrinsic $\gamma^\mu$ part but not from
the model-dependent part with $D=D_{\rm cov}$ or $D_{\rm LF}$ factor
as we discussed in the previous section.

Since the frame-independent (or covariant) form factors
$(g^{\rm DY}, a^{\rm DY}_+,f^{\rm DY})$ have already been given
in our previous analysis~\cite{ZM05}, we
do not list them here again. Including the zero-mode contribution given by Eq.~(\ref{JPZM2}),
we now obtain the form factor $a^{\rm DY}_-(q^2)$ as follows:

\bea\label{Fam}
a^{\rm DY}_-(q^2)&=&\frac{N}{8\pi^3}
\int^1_0
\frac{dx}{(1-x)}\int d^2{\bf k}_\perp\;
\chi_1\chi_2
\biggl\{ (2x -3){\cal A}_1
+ \frac{{\bf k}_\perp\cdot{\bf q}_\perp}{q^2}[ (7-6x) m_1
\nonumber\\
&&- m_2 - (4-6x) m]
+\frac{2}{q^2}(m_1 - m)\biggl[x Z_2 - 2\frac{({\bf k}_\perp\cdot{\bf q}_\perp)^2}{q^2}\biggr]
\nonumber\\
&&-\frac{2}{(1-x)D}\biggr(
({\bf k}_\perp\cdot{\bf k}'_\perp + {\cal A}_1{\cal B}_2)
\biggl[ (1-x) + \frac{Z_2}{q^2} - \frac{{\bf k}_\perp\cdot{\bf q}_\perp}{q^2} \biggr]
\nonumber\\
&&-
\biggl[ (1-x) Z_2 + 2{\bf k}^2_\perp + 2m{\cal A}_1
-2 (1-x)[ M^2_2 - q^2
\nonumber\\
&&+ (m_2 + m)(m_1-m) ]\frac{{\bf k}_\perp\cdot{\bf q}_\perp}{q^2}
\biggr]\biggl[ (1-x) - \frac{{\bf k}_\perp\cdot{\bf q}_\perp}{q^2} \biggr]
\biggr)
\biggr\},
\eea
where ${\cal A}_i = (1-x) m_i + xm(i=1,2)$ and ${\cal B}_2 = xm -(1-x) m_2$.

\begin{table}
\caption{The existence ($\rm O (\rm source\; element)$) or absence ($\rm X$) of the zero-mode contribution to the weak
form factors for the semileptonic $P\to V\ell\nu_{\ell}$ decays depending on the current matrix
element $\la J^\mu_{V-A}\ra_h$ and the vector meson vertex
$\Gamma^\mu=\gamma^\mu - (P_2-2k)^\mu/D$ with various $D$ factors such as
$D_{\rm con}\sim (1/x)^0$, $D_{\rm cov}\sim (1/x)^{1}$,
and $D_{\rm LF}\sim (1/x)^{1/2}$ as $x\to 0$.}
\label{t1}
\renewcommand{\tabcolsep}{1.0pc} % enlarge column spacing
\begin{center}
\begin{tabular}{@{}lllll} \hline
 & $g$ & $a_+$ & $a_-$ & $f$  \\
\hline
 & $\la J^+_{V}\ra_1$ & $\la J^+_{A}\ra_1$
& ($\la J^+_{A}\ra_0$, $\la J^+_{A}\ra_1$,$\la J^\perp_{A}\ra_1$)
& ($\la J^+_{A}\ra_0$, $\la J^+_{A}\ra_1$)\\
\hline
$\gamma^\mu$ & $\rm X$ & $\rm X$ & $\rm O$ ($\la J^\perp_{A}\ra^{\rm Z.M.}_1$)& $\rm X$  \\
$\frac{(P_2-2k)^\mu}{D_{\rm con}}$ & $\rm X$
& $\rm X$ & $\rm O$($\la J^+_{A}\ra^{\rm Z.M.}_0,\la J^\perp_{A}\ra^{\rm Z.M.}_1$)
& $\rm O$ ($\la J^+_{A}\ra^{\rm Z.M.}_0$) \\
$\frac{(P_2-2k)^\mu}{D_{\rm cov}}$ & $\rm X$ & $\rm X$ & $\rm X$ & $\rm X$  \\
$\frac{(P_2-2k)^\mu}{D_{\rm LF}}$ & $\rm X$ & $\rm X$ & $\rm X$ & $\rm X$ \\
\hline
\end{tabular}
\end{center}
\end{table}

In Table~\ref{t1}, we summarize our findings on the existence/absence of the zero-mode
contribution to the hadronic form factors ($g, a_\pm, f$)
for the semileptonic $P\to V\ell\nu_{\ell}$ decays
depending on the current matrix
element $\la J^\mu_{V-A}\ra_h$ and the vector meson vertex
$\Gamma^\mu=\gamma^\mu - (P_2-2k)^\mu/D$ with various $D$ factors.
Since our findings on the existence/absence of the zero mode are based on the method of power counting, our conclusion applies to other methods of regularization as far as the regularization doesn't change the power counting in the form factor calculation. For example, as discussed by Jaus in Ref.~\cite{Ja99}, some other multipole type ansatz in the method of regularization wouldn't change the conclusion drawn by the monopole type ansatz.

We should note, however, that Jaus's prescription~\cite{Ja99} is
valid only for the case of $D=D_{\rm con}$  but not for
the more realistic $D=D_{\rm cov}$ or $D_{\rm LF}$ case. Essentially, Jaus's
prescription corresponds to the replacement
$p^-_1/D\to -Z_2/D$ for the vector term and
$p^-_1{\bf p}_{1\perp}/D\to -({\bf q}_\perp/D)[A^{(1)}_2 Z_2 + (q\cdot P/q^2) A^{(2)}_1]$ for the
tensor term regardless of the $D$ factor used~\cite{Bc09}.
Indeed he applied this prescription to the more realistic $D=D_{\rm LF}$ factor case~\cite{Ja99}.
However, we show that such prescription is valid only for $D=D_{\rm con}$ but not
for $D=D_{\rm cov}$ or $D_{\rm LF}$.
For $D=D_{\rm cov}$ or $D_{\rm LF}$, the valid replacement should be
$p^-_1/D\to 0$ and $p^-_1{\bf p}_{1\perp}/D\to 0$.

In terms of current matrix elements, $\la J^+_A\ra^{\rm Z.M.}_0$
comes only from $p^-_1/D$ term but $\la J^\perp_A\ra^{\rm Z.M.}_0$
comes from both $p^-_1$ and $p^-_1{\bf p}_{1\perp}/D$ terms.
We thus stress that $\la J^+_A\ra^{\rm Z.M.}_0$ is absent
and the form factor $f(q^2)$ is immune to the zero mode when
$D=D_{\rm cov}$ or $D_{\rm LF}$ is used.
Although the form factor $a_-(q^2)$ receives the zero-mode contribution from
$\la J^\perp_A\ra^{\rm Z.M.}_0$, it comes only from the intrinsic
$\gamma^\mu$ part (i.e. $p^-_1\to -Z_2$) but not from the
$(P_2-2k)^\mu/D$ part (i.e. $p^-_1{\bf p}_{1\perp}/D\to 0$ for $D=D_{\rm cov}$ or $D_{\rm LF}$).
Such absence of the zero mode, i.e. $p^-_1/D\to 0$ and
$p^-_1{\bf p}_{1\perp}/D\to 0$ are not realized in Jaus's approach~\cite{Ja99} for $D=D_{\rm cov}$ or
$D_{\rm LF}$.

For an explicit demonstration of our findings, we performed the numerical calculation
of the $B\to\rho$ transition form factors
using the model parameters for $B$ and $\rho$ used in Refs.~\cite{BCJ03,ZM05}.
The frame-independence of our results was also checked by comparing the results from the $q^+=0$ frame
with those from the $q^+>0$ frame which is summarized in the appendix B.
Through the manifestly covariant calculation with $D=D_{\rm con}$ and $D_{\rm cov}$,
we indeed confirmed that our findings of the zero-mode contributions are correct.

\begin{figure}
\begin{center}
\includegraphics[width=2.5in,height=2.5in]{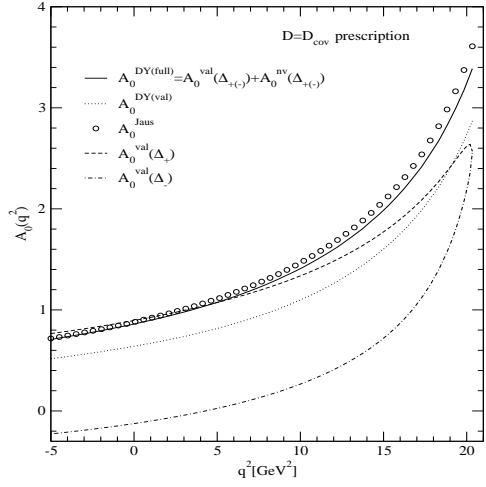}
\caption{The weak form factor $A_0(q^2)$ for
$B\to\rho$ transition for the vector meson vertex
$\Gamma^\mu=\gamma^\mu-(P_2-2k)^\mu/D_{\rm cov}$.}
\label{fig2}
\end{center}
\end{figure}

In Fig.~\ref{fig2}, we present the form factor $A_0(q^2)$ for the vector meson vertex
$\Gamma^\mu=\gamma^\mu-(P_2-2k)^\mu/D_{\rm cov}$.
The solid ($A_{0}^{\rm DY(full)}$) and dotted ($A_{0}^{\rm DY(val)}$)
lines represent the full (i.e. valence + zero-mode) result
and the valence contribution in the $q^+=0$ frame, respectively.
That is, the difference between the two results (i.e.
$A_{0}^{\rm DY(full)}-A_{0}^{\rm DY(val)}$)
represents the zero-mode contribution $A_{0}^{\rm Z.M.}$ to $A_0(q^2)$.
The circle ($A^{\rm Jaus}_0$) represents the result obtained from the Jaus's prescription~\cite{Ja99}.
The dashed and dot-dashed lines represent the valence
results obtained from the purely longitudinal $q^+>0$ frame with
$\Delta_+$ and $\Delta_-$ given by Eq.~(\ref{pl:2}), respectively.
We note that the valence contribution in the $q^+>0$ frame depends on the
direction of the daughter meson recoiling in the positive ($\Delta_+$) or
the negative ($\Delta_-$) $z$-direction relative to the parent meson.
Including the nonvalence contribution, however, the full result in the $q^+>0$ frame
is in complete agreement with $A^{\rm DY(full)}_0$ in the $q^+=0$ frame.
On the other hand, $A^{\rm Jaus}_0$ shows a small but clear deviation from our full result.
Since there is no zero-mode contribution from the
$p^-_1/D_{\rm cov}$ term, the zero-mode contribution included in the $D_{\rm con}$ case is
absent in the $D_{\rm cov}$ case. Such absence of the zero mode is not
realized in Jaus's approach~\cite{Ja99}.
The numerical deviation between our result (solid line) and Jaus's result (circle) shown
in Fig.~\ref{fig2} is due to this difference.

For the $D=D_{\rm LF}$ case, although we do not know how to
compute the nonvalence diagram, we can still use our counting rule for the longitudinal momentum
fraction factors to check the existence of the zero mode.
As summarized in Table~\ref{t1}, the zero-mode contributions from
$p^-_1/D_{\rm LF}$ and $p^-_1{\bf p}_{1\perp}/D_{\rm LF}$
do not exist as in the case of $D_{\rm cov}$.

\section{Summary and Discussion}
In this work, we have analyzed the zero-mode contribution to the weak
transition form factors between
pseudoscalar and vector mesons. For the phenomenologically accessible vector
meson vertex $\Gamma^\mu=\gamma^\mu - (P_2-k)^\mu/D$, we discussed the three
typical cases of the $D$ factor which may be classified by the differences
in the power counting of the LF energy (or longitudinal momentum fraction $x$)
$p^-_1\sim 1/x$, i.e.:
(1) $D_{\rm con}=M_V + m_2 + m\sim (1/x)^0$,
(2) $D_{\rm cov}=[2k\cdot P_2 + M_2(m_2 + m) -i\epsilon]/M_2 \sim (1/x)^1$,
and
(3) $D_{\rm LF}=M'_0 + m_2 + m\sim (1/x)^{1/2}$.
Our main idea to obtain the weak transition form factors is first to find if the
zero-mode contribution exists or not for the given form factor using the power counting
method. If it exists, then the separation of the on-mass-shell propagating part from
the off-mass-shell instantaneous part is useful since the latter is responsible for the
zero-mode contribution.

Our findings on the existence/absence of the zero-mode contribution to the weak transition
form factors $(g,a_\pm, f)$ are summarized in Table~\ref{t1}.
We found that the form factors $g(q^2)$ and $a_+(q^2)$ are immune
to the zero-mode contribution in all three cases of the $D$ factors.
However, the existence/absence of
the zero mode in the form factors $a_-(q^2)$ and $f(q^2)$ depends on the nature of the
$D$ factors. For the form factor $f(q^2)$, while the zero-mode contribution exists in the
$D_{\rm con}$ case, the other two cases such as $D_{\rm cov}$
and $D_{\rm LF}$ are immune to the zero-mode contribution.
We also should note that the zero-mode contribution to $f(q^2)$ does not exist in the simple vector meson
vertex $\Gamma^\mu=\gamma^\mu$ (i.e. $1/D=0$ case) as we have already shown in~\cite{BCJ03}.
For the form factor $a_-(q^2)$, however, the zero-mode contribution exists
in the case of $1/D=0$.
Including the $D$ factor, the zero-mode contribution coming from the $D$ factor
exists in the $D_{\rm con}$ case but not in the other two cases of $D_{\rm cov}$
and $D_{\rm LF}$. That is, if one uses the $D_{\rm cov}$
or $D_{\rm LF}$ in the phenomenological vector meson vertex, our results show that
the form factors $g(q^2)$, $a_+(q^2)$, and $f(q^2)$ are immune to the zero-mode contribution,
but the form factor $a_-(q^2)$ receives the zero mode coming only from the simple
vertex $\gamma^\mu$ term. We also found the corresponding zero-mode operator for $a_-(q^2)$ that
is convoluted with the initial and final state LF wave functions (see Eq.~(\ref{JPZM2})).
This provides a well-established basis of LF approach to compute the weak transition form
factors between pseudoscalar and vector mesons without missing any zero-mode contribution.
The covariance (i.e., frame independence) of our model for the
cases of $D=D_{\rm con}$ and $D_{\rm cov}$ has been
checked by performing the LF calculation in the $q^+=0$ frame in parallel with the purely
longitudinal $q^+> 0$ frame using the exactly solvable covariant fermion field theory
model in $(3+1)$ dimensions.

All of these findings stem from the fact that the zero-mode contribution from the $D$ factor
is absent if the denominator $D$ of the vector meson vertex
$\Gamma^\mu=\gamma^\mu - (P_2-k)^\mu/D$
contains the term proportional to the LF energy (or longitudinal momentum fraction $x$)
$(p^-_1)^n\sim (1/x)^n$ with the power $n>0$.
While the correct implementation of zero-mode contributions cannot
solve all the problems in the phenomenology, at least the Lorentz covariance of the result
can be assured in the LFQM. This certainly benefits the hadron phenomenology.

\section*{Acknowledgment}
 The work of H.-M.Choi was supported by the Korea
Research Foundation Grant funded by the Korean
Government(KRF-2010-0009019) and that of C.-R.Ji by the U.S.
Department of Energy(No. DE-FG02-03ER41260).

\appendix
\section{Trace terms $(S^\mu_V)_h$ and
$(S^\mu_A)_h$ in Eq.~(\ref{eqq:4})}
In this appendix we summarize the LF results of the trace terms $[S^\mu_{V(A)}]_h$
in Eq.~(\ref{eqq:4}) by separating it
into the on-mass-shell propagating part $[S^\mu_{V(A)}]^{\rm on}_h$ and
the off-mass-shell instantaneous part $[S^\mu_{V(A)}]^{\rm inst}_h$ as follows
\bea\label{SVon1}
(S^\mu_V)_h^{\rm on} &=& 4i\vep^{\mu\nu\rho\sigma}\biggl\{
[ m_1 p_{2\rm on\nu} k_{\rm on\rho}
 - m_2 p_{1\rm on\nu} k_{\rm on\rho}
 - m p_{2\rm on\nu} p_{2\rm on\rho} ]\ep^*_\sigma
 \nonumber\\
 &&+ \frac{2k_{\rm on}\cdot\ep^*}{D_{\rm on}}
 p_{1\rm on\nu} p_{2\rm on\rho} k_{\rm on\sigma}\biggr\},
 \eea
 \bea\label{SVon2}
 (S^\mu_V)_h^{\rm inst} &=&
 2i\vep^{\mu\nu+\sigma}\biggl\{
 [ \delta p_1^-(m p_{2\rm on} + m_2 k_{\rm on})_\nu
 - \delta p_2^-(m p_{1\rm on} + m_1 k_{\rm on})_\nu
 \nonumber\\
 &&- \delta k^-(m_2 p_{1\rm on} - m_1 p_{2\rm on})_\nu ]\ep^*_\sigma
 -\frac{2k_{\rm on}\cdot\ep^*
 + \ep^{*+}\delta k^-}{D_{\rm on}+\delta D}
 \nonumber\\
 &&\times
 [  p_{2\rm on\nu} k_{\rm on\sigma}\delta p_1^-
 -  p_{1\rm on\nu} k_{\rm on\sigma} \delta p_2^-
 +  p_{1\rm on\nu} p_{2\rm on\sigma} \delta k^- ]\biggr\}
 \nonumber\\
 &&
 + 4i\vep^{\mu\nu\rho\sigma}\frac{\delta k^-}{D_{\rm on}+\delta D}
 \ep^{*+}p_{1\rm on\nu} p_{2\rm on\rho} k_{\rm on\sigma},
\eea
and
\bea\label{SAon1}
(S^\mu_A)_h^{\rm on}&=& 4m_1[ (k_{\rm on}\cdot\ep^*)p_{2\rm on}^\mu
+ (p_{2\rm on}\cdot\ep^*)k_{\rm on}^\mu
- (k_{\rm on}\cdot p_{2\rm on})\ep^{*\mu}]
\nonumber\\
&&
- 4m_2[ (k_{\rm on}\cdot\ep^*)p_{1\rm on}^\mu
+ (p_{1\rm on}\cdot\ep^*)k_{\rm on}^\mu
+ (k_{\rm on}\cdot p_{1\rm on})\ep^{*\mu}]
\nonumber\\
&&
 + 4m[ (p_{2\rm on}\cdot\ep^*)p_{1\rm on}^\mu
+ (p_{1\rm on}\cdot\ep^*)p_{2\rm on}^\mu
- (p_{1\rm on}\cdot p_{2\rm on})\ep^{*\mu}]
\nonumber\\
&& + 4m m_1 m_2\ep^{*\mu}
 - 8\frac{k_{\rm on}\cdot\ep^*}{D_{\rm on}}
[ (p_{2\rm on}\cdot k_{\rm on} - m_2 m)p_{1\rm on}^\mu
\nonumber\\
&&+ (p_{1\rm on}\cdot k_{\rm on} + m_1 m)p_{2\rm on}^\mu
- (p_{1\rm on}\cdot p_{2\rm on} + m_1 m_2)k_{\rm on}^\mu ],
\eea
\bea\label{SAon2}
(S^\mu_A)_h^{\rm inst}&=& -2\delta k^-\biggl\{
m_1 [p_{2}^+\ep^{*\mu} - p_{2\rm on}^\mu\ep^{*+} - (p_{2\rm on}\cdot\ep^*)g^{\mu+}]
+ m_2 [p_{1}^+\ep^{*\mu}
\nonumber\\
&&+ p_{1\rm on}^\mu\ep^{*+} - (p_{1\rm on}\cdot\ep^*)g^{\mu+}]
+\frac{2k_{\rm on}\cdot\ep^* + \ep^{*+}\delta k^-}{D_{\rm on} + \delta D}
[ p_{1}^+ p_{2\rm on}^\mu
\nonumber\\
&&+ p_{1\rm on}^\mu p_{2}^+
- (p_{1\rm on}\cdot p_{2\rm on} + m_1 m_2)g^{\mu+}] \biggr\}
\nonumber\\
&& -2\delta p_1^-\biggl\{ m_2 [k^+\ep^{*\mu}
- k_{\rm on}^\mu\ep^{*+} + (k_{\rm on}\cdot\ep^*)g^{\mu+}]
+ m [p_{2}^+\ep^{*\mu}
\nonumber\\
&& - p_{2\rm on}^\mu\ep^{*+}
- (p_{2\rm on}\cdot\ep^*)g^{\mu+}]
 +\frac{2k_{\rm on}\cdot\ep^* + \ep^{*+}\delta k^-}{D_{\rm on} + \delta D}
[ p_{2\rm on}^\mu k^+
\nonumber\\
&&- p_{2}^+ k_{\rm on}^\mu
+ (p_{2\rm on}\cdot k_{\rm on} - m_2 m)g^{\mu+}] \biggr\}
\nonumber\\
&&-2\delta p_2^-\biggl\{
m_1 [k^+\ep^{*\mu} - k_{\rm on}^\mu\ep^{*+} - (k_{\rm on}\cdot\ep^*)g^{\mu+}]
+ m [p_{1}^+\ep^{*\mu}
\nonumber\\
&&- p_{1\rm on}^\mu\ep^{*+} - (p_{1\rm on}\cdot\ep^*)g^{\mu+}]
 +\frac{2k_{\rm on}\cdot\ep^* + \ep^{*+}\delta k^-}{D_{\rm on} + \delta D}
[ p_{1\rm on}^\mu k^+
\nonumber\\
&& - p_{1}^+ k_{\rm on}^\mu + (p_{1\rm on}\cdot k_{\rm on} - m_1 m)g^{\mu+}] \biggr\}
+2m_1 \ep^{*+}g^{\mu+}\delta p_2^- \delta k^-
\nonumber\\
&&
 + 2g^{\mu+} \delta p_1^- \delta p_2^-
\biggl\{ m\ep^{*+} - \frac{2k_{\rm on}\cdot\ep^* + \ep^{*+}\delta k^-}{D_{\rm on} + \delta D}k^+
\biggr\}
\nonumber\\
&&-4\frac{\delta k^-}{D_{\rm on} +\delta D}\ep^{*+}
[ (p_{2\rm on}\cdot k_{\rm on} - m_2 m)p_{1\rm on}^\mu
\nonumber\\
&&+ (p_{1\rm on}\cdot k_{\rm on} + m_1 m)p_{2\rm on}^\mu
- (p_{1\rm on}\cdot p_{2\rm on} + m_1 m_2)k_{\rm on}^\mu ],
\eea

where $D_{\rm on}$ is the denominator factor $D$ when $k=k_{\rm on}$
and $\delta D$ is the difference between $D$ and $D_{\rm on}$, i.e.
$\delta D=D(k) - D_{\rm on}(k_{\rm on})$. For the
$D_{\rm con(LF)}$ factor, $\delta D_{\rm con (LF)} =0$ and
$D_{\rm on}=D_{\rm con(LF)}$.
For the $D_{\rm cov}$ factor including the four momentum
$k$ explicitly, however, one obtains
$\delta D_{\rm cov} =\delta k^-(P^+_2/M_2)$.
In Eq.~(\ref{SAon2}), we have also used $P_2\cdot\ep^*=0$.

\section{Form factors in the purely
longitudinal frame}
The purpose of this appendix is to show the frame-independence (i.e. covariance) of our
result obtained from the $q^+=0$ frame by comparing with the result obtained from the
$q^+>0$ frame, which is summarized in this appendix.

In the reference frame where $q^+ > 0$ and ${\bf P}_{1\perp}=0$, the
(timelike) momentum transfer $q^2=(P_1-P_2)^2$ is given by
\begin{equation}
{\label{tq2}}
q^2=q^+q^- - {\bf q}^2_\perp =\Delta\biggl(M^2_1 -
\frac{M^2_2}{1-\Delta}\biggr) -\frac{{\bf q}^2_\perp}{1-\Delta},
\end{equation}
where $q^+=\Delta P^+_1$.
In this frame, only the plus component of the $V-A$ current can be utilized
for the calculations of
LF valence[Fig.~\ref{fig1}(b)] and nonvalence [Fig.~\ref{fig1}(c)] diagrams.

In the valence region $0<k^+<P^+_2$ (i.e. $\Delta<x<1$), the
pole at $k^-=k^-_{\rm on}$ (from the spectator quark) is located in the
lower half plane of the complex $k^-$ variable.
Thus, the Cauchy integration formula for the $k^-$-integral in
Eq.~(\ref{eq:4}) yields
\be{\label{jval}}
\la J^{\mu}_{h}\ra^{\rm val} =
\frac{N}{16\pi^3}\int^1_\Delta
\frac{dx}{1-x}\int d^2{\bf k}_\perp
\chi_1(x,{\bf k}_\perp)S^\mu_h(k^-_{\rm on})
\chi_2(x',{\bf k'}_\perp),
\ee
where $x'=(x-\Delta)/(1-\Delta)$ and
${\bf k'}_\perp={\bf k}_\perp + (1-x'){\bf q}_\perp$.

In the nonvalence region $P^+_2<k^+<P^+_1$ (i.e. $0<x<\Delta$)
the poles at $p^-_1=p^-_{1\rm on}(m_1)$ (from the struck quark propagator)
and $p^-_1=p^-_{1\rm on}(\Lambda_1)$ (from the smeared quark-gauge-boson vertex) are
located in the upper half plane of the complex $k^-$ variable.
Thus, the Cauchy integration over $k^-$ in Eq.~(\ref{eq:4})
yields
\bea{\label{jnv}}
\la J^\mu_{h}\ra^{\rm nv}
&=&
\frac{N}
{16\pi^3(\Lambda^2_1-m^2_1)}
\int^\Delta_0\frac{dx}{(1-x)(\Delta-x)x''(1-x'')}
\nonumber\\
&&\times
\int d^2{\bf k}_\perp
\biggl\{
\frac{S^\mu_{h}(p^-_{1\rm on}(\Lambda_1))}
{(M^2_1-M^2_{\Lambda_1})(q^2-M^2_{\Lambda_1\Lambda_2})
(q^2-M^2_{\Lambda_1m_2})}
\nonumber\\
&&\;\;\;\;
- \frac{S^\mu_{h}(p^-_{1\rm on}(m_1))}
{(M^2_1-M^2_{0})(q^2-M^2_{m_1\Lambda_2})(q^2-M^2_{m_1m_2}))}
\biggr\},
\eea
where
\bea
M^2_{ab}&=&
\frac{{\bf k''}^2_\perp+a^2}{x''}
+\frac{{\bf k''}^2_\perp + b^2}{1-x''},
\eea
and
\bea
x''&=& \frac{x}{\Delta}, \;
{\bf k''}_\perp = {\bf k}_\perp + x''{\bf q}_\perp.
\eea
The explicit forms of the trace terms
$(S^+_{V-A})^{\rm nv}_h(p^-_{1\rm on}(\Lambda_1))$ for the vector and
axial-vector currents are given by

\bea\label{Snvh}
(S^+_V)^{\rm nv}_{h=1}
 &=&
-\frac{2}{\sqrt{2}}\varepsilon^{+-xy}\biggl\{
q^L{\cal A}_1
+ k^L [m_1-m_2 - \Delta(m_1-m)]
\nonumber\\
&& +
\frac{2}{D}[{\bf k}^2_\perp q^L - ({\bf k}_\perp\cdot{\bf q}_\perp)k^L]\biggl\},
\nonumber\\
(S^+_A)^{\rm nv}_{h=1}
&=& \frac{4}{\sqrt{2}}\biggl\{
(2x'-1)q^L{\cal A}_1
+ k^L [ (2x-1-\Delta)(m_1-m)  - m_2 -m]
\nonumber\\
&&
-\frac{2( (1-x')q^L + k^L)}{(1-x')D}
[{\bf k}_\perp\cdot{\bf k'}_\perp
+ {\cal A}_1 {\cal B}'_2
\nonumber\\
&&+xx'(1-x)(M^2_1-M^2_{\Lambda_1}) ]
\biggr\},
\nonumber\\
(S^+_A)^{\rm nv}_{h=0}
&=&-\frac{4}{(1-x')M_2}\biggl\{
{\cal A}_1 [x'(1-x')M^2_2 + m_2 m - (1-x')^2 q^2 ]
\nonumber\\
&&+ {\bf k}^2_\perp ( {\cal A}_1 + m_2 - m)
+ (1-x'){\bf k}_\perp\cdot{\bf q}_\perp ( 2{\cal A}_1 + m_2 - m )
\nonumber\\
&&
+ x(1-x)m_2 (M^2_1 - M^2_{\Lambda_1})
-\frac{1}{xD}[ x(1-x')M^2_2 - (1-\Delta) x M^2_1
\nonumber\\
&&
+ (1-\Delta)(\Lambda^2_1 + {\bf k}^2_\perp) + x(1-x')q^2
- 2x{\bf k}_\perp\cdot{\bf q}_\perp ]
[ {\bf k}_\perp\cdot{\bf k'}_\perp
+ {\cal A}_1 {\cal B}'_2
\nonumber\\
&&+ xx'(1-x)(M^2_1 -M^2_{\Lambda_1})]
\biggr\},
\eea

where ${\cal A}_i = (1-x) m_i + xm(i=1,2)$ and ${\cal B}'_2=x'm-(1-x')m_2$.
The trace terms $S^+_h(p^-_{1\rm on}(m_1))$ for the vector and axial-vector currents
can be obtained by the replacement $\Lambda_1\to m_1$ in Eq.~(\ref{Snvh}).
We also note that the trace terms
$S^+_h(p^-_{1\rm on}(\Lambda_1))$  and $S^+_h(p^-_{1\rm on}(m_1))$ in the
nonvalence region include both the on-mass-shell quark propagating part
and the off-mass-shell instantaneous part.

The relations between the current matrix elements and
the weak form factors in this $q^+ > 0$ frame are as follows:
\bea{\label{ch:4a}}
\la J^+_V\ra_{h=1} &=&
-\frac{1}{\sqrt{2}}\varepsilon^{+-xy}q^L g(q^2),
\eea
for the vector current and
\bea{\label{ch:5a}}
\la J^+_A\ra_{h=1}
&=& \frac{q^L}{(1-\Delta)\sqrt{2}}
\biggl[ (2-\Delta) a_+(q^2)
+ \Delta a_-(q^2)\biggr],
\nonumber\\
\la J^+_A\ra_{h=0}
&=& \frac{1-\Delta}{2M_2}\biggl\{ 2f(q^2) +
\biggl[ M^2_1 - \frac{M^2_2}{(1-\Delta)^2}
\nonumber\\
&&- \frac{q^2}{(1-\Delta)^2}\biggr]
[ (2-\Delta)a_+(q^2) + \Delta a_-(q^2)]
\biggr\},
\eea
for the axial-vector current.
In the purely longitudinal momentum
$q^+ > 0$ and ${\bf q}_\perp=0$ frame, where
\bea{\label{pl:1}}
q^2&=&q^+q^- =\Delta\biggl(M^2_1 - \frac{M^2_2}{1-\Delta}\biggr),
\eea
there are two solutions of $\Delta$ for a given $q^2$, i.e.,
\be{\label{pl:2}}
\Delta_{\pm}=
\frac{M^2_1 - M^2_2+q^2 \mp
\sqrt{ (M^2_1 - M^2_2+q^2)^2-4M^2_1q^2}}
{2M^2_1},
\ee
where the $+(-)$ sign in Eq.~(\ref{pl:2}) corresponds to the daughter
meson recoiling in the positive(negative) $z$-direction relative to
the parent meson.

At zero recoil ($q^2=q^2_{\rm max}$) and maximum
recoil ($q^2=0$), $\Delta_\pm$ are given by
\bea{\label{apm}}
\Delta_+(q^2_{\rm max})&=&\Delta_-(q^2_{\rm max})=1-\frac{M_2}{M_1},
\nonumber\\
\Delta_+(0)&=&0,\;\;
\Delta_-(0)=1-\biggl(\frac{M_2}{M_1}\biggr)^2.
\eea

The form factors should in principle be independent of the recoil directions
($\Delta_\pm$) if the nonvalence contributions are added to the valence
ones.
While the form factor $g(q^2)$ in the $q^+>0$ frame can be obtained
directly from Eq.~(\ref{ch:4}), the form factor $f(q^2)$ can be
obtained only after $a_\pm(q^2)$ are calculated.

To illustrate this, we define
\bea{\label{pl:3}}
\la J^+_A\ra_{h=1} |_{\Delta=\Delta_\pm}
&\equiv& \frac{q^L}{\sqrt{2}}I^+_A(\Delta_\pm).
\eea
Then we obtain from Eq.~(\ref{ch:5a})
\bea{\label{pl:4}}
a_+(q^2)&=&\frac{1}{2(\Delta_- -\Delta_+)}[
\Delta_-(1-\Delta_+)I^+_A(\Delta_+)
- \Delta_+(1-\Delta_-) I^+_A(\Delta_-)],
\nonumber\\
a_-(q^2)&=&\frac{1}{2(\Delta_+ -\Delta_-)}[
(1+\Delta_+)(2-\Delta_-)I^+_A(\Delta_+)
\nonumber\\
&&- (1-\Delta_-)(2-\Delta_+) I^+_A(\Delta_-)],
\eea
and
\be{\label{pl:5}}
f(q^2)= \frac{M_2}{1-\Delta}\la J^+_A\ra_{h=0}
-\frac{1}{2}\biggl(M^2_1-\frac{M^2_2}{(1-\Delta)^2}\biggr)
\biggl[(2-\Delta)a_+(q^2) + \Delta a_-(q^2) \biggr],
\ee
where $\Delta$ in Eq.~(\ref{pl:5}) can be either $\Delta_+$ or
$\Delta_-$.
As one can see from Eqs. (\ref{ch:4a}) and (\ref{ch:5a}),
one should be careful in setting ${\bf q}_\perp = 0$ to get the correct
results in the purely longitudinal frame.  One cannot simply set
${\bf q}_\perp=0$ from the start, but may set it to zero only after the
form factors are extracted.

\end{document}